\documentclass{aa}
\usepackage{graphicx}
\usepackage{txfonts}
\begin{document}
  \title{Extrasolar planets and brown dwarfs around A-F type stars
    \thanks{Based on observations made with the {\small ELODIE}
    spectrograph at the Observatoire de Haute-Provence (CNRS,
    France).}  } 

  \subtitle{IV. A candidate brown dwarf around the A9V
    pulsating star HD\,180777.}
  
  \author{
    F. Galland \inst{1,2}
    \and
    A.-M. Lagrange \inst{1}
    \and
    S. Udry \inst{2}
   \and
    J.-L. Beuzit \inst{1}
     \and
    F. Pepe \inst{2}
    \and
    M. Mayor \inst{2}
  }
  
  \offprints{
    F. Galland,\\
    \email{Franck.Galland@obs.ujf-grenoble.fr}
  }

  \institute{
    Laboratoire d'Astrophysique de l'Observatoire de Grenoble,
    Universit\'e Joseph Fourier, BP 53, 38041 Grenoble, France
    \and
    Observatoire de Gen\`eve, 51 Ch. des Maillettes, 1290 Sauverny, Switzerland
  }

  \date{Received August 22, 2005 / Accepted February 06, 2006}

  \abstract{

    We present here the detection of a brown dwarf orbiting the A9V
    star HD\,180777. The radial velocity measurements, obtained with
    the {\small ELODIE} echelle spectrograph at the Haute-Provence
    Observatory, show a main variation with a period of
    28.4~days. Assuming a primary mass of 1.7~M$_{\odot}$, the best
    Keplerian fit to the data leads to a minimum mass of
    25~M$_{\rm{Jup}}$ for the companion (the true mass could be
    significantly higher).  We also show that, after substraction of
    the Keplerian solution from the radial velocity measurements, the
    residual radial velocities are related to phenomena intrinsic to
    the star, namely pulsations with typical periods of $\gamma$~Dor
    stars. These results show that in some cases, it is possible to
    disentangle radial velocity variations due to a low mass companion
    from variations intrinsic to the observed star.

  }
  
  \maketitle
  
  \section{Introduction}

  Radial velocity surveys have lead to the detection of more than 160
  planets during the past decade~\footnote{A comprehensive list of
  known exoplanets is available at
  http://www.obspm.fr/encycl/cat1.html}.  We are performing a radial
  velocity survey dedicated to the search for extrasolar planets and
  brown dwarfs around a volume-limited sample of more massive stars
  than currently done, namely A-F main-sequence stars (i) with the
  {\small ELODIE} fiber-fed echelle spectrograph (\cite{Baranne96})
  mounted on the 1.93-m telescope at the Observatoire de
  Haute-Provence (CNRS, France) in the northern hemisphere, and (ii)
  with the {\small HARPS} spectrograph (\cite{Pepe02}) installed on
  the 3.6-m ESO telescope at La Silla Observatory (ESO, Chile) in the
  southern hemisphere.  Finding planets and brown dwarfs around
  massive stars is important, as this will allow us to test planetary
  formation and evolution processes around a wider variety of objects.

  As A-F main-sequence stars exhibit a small number of stellar lines,
  usually broadened and blended by stellar rotation, we developed a
  new radial velocity method that is described in \cite{Galland05a}
  (Paper\,I), together with the detection limits we achieved and the
  estimates of the minimum detectable masses. The first results of the
  survey are (i) discovering with {\small ELODIE} a planet around an
  F6V star (\cite{Galland05b}, Paper\,II), and (ii) finding the limits
  to the presence of an inner giant planet around $\beta$ Pictoris, with
  {\small HARPS} and {\small CORALIE} (\cite{Galland06a}, Paper\,III);
  in this last case, the observed radial velocity variations are
  attributed to $\delta$~Scuti type pulsations.
  
  We present here the detection of a brown dwarf around one of the
  objects surveyed with {\small ELODIE}, HD\,180777.  Section 2
  provides the stellar properties of this star.  Section 3 explains
  the measurement of the radial velocities, and the Keplerian solution
  associated to the main radial velocity variations is derived in
  Sect.~4.  In Sect.~5, we rule out other possible origins of these
  main radial velocity variations.  The large radial velocity
  residuals around the orbital solution are interpreted in terms of
  pulsations in Sect.~6.

  \section{Stellar properties}

  HD\,180777 (HIP\,94083, HR\,7312) is located at 27.3~pc from the Sun
  (ESA 1997).  Its projected rotational velocity, calculated using
  auto-correlation, is $v\sin{i}$\,=\,50\,km\,s$^{\rm -1}$; if the
  true rotational velocity of this star meets the mean rotational
  velocity of A9V type stars, namely 125\,km\,s$^{\rm -1}$
  (\cite{Gray05}), the value of $\sin{i}$ would be 0.4\,. The
  effective temperature $T_{eff}$\,=\,7250\,K and the surface gravity
  $\log{g}$\,=\,4.34 are taken from \cite{King03}. The stellar
  properties are summarized in
  Table\,\ref{Table_hd180777_stelpar}. Note that the MK spectral type
  of HD\,180777 varies from A9V to F2V depending on the authors; yet,
  the spectral type A9V is more frequent (see e.g. the HIPPARCOS
  catalogue, ESA 1997, or the Bright Star Catalogue,
  \cite{Hoff91}). Given this spectral type, we deduce a stellar mass
  of 1.7$\pm$0.1~M$_\odot$.

  HD\,180777 belongs to the range of B-V where the instability strip
  intersects with the main sequence. This region contains the
  pulsating $\delta$~Scuti (\cite{Handler02}, \cite{Breger00}) and
  $\gamma$~Dor stars (\cite{Mathias04}), with respective stellar mass
  ranges of [1.5, 2.2] M$_\odot$ and [1.2, 1.9] M$_\odot$.  With a
  mass of $\approx$ 1.7 M$_\odot$, HD\,180777 could then belong to one
  or the other class of pulsating stars. We show in Sect.~6 that
  HD\,180777 actually undergoes pulsations with frequencies associated
  to the $\gamma$~Dor stars.

  \begin{table}[t!]
    \caption{HD\,180777 stellar properties. Photometric and
      astrometric data are extracted from the HIPPARCOS catalogue
    (ESA 1997). Spectroscopic data are from \cite{King03}.
    }
    \label{Table_hd180777_stelpar}
    \begin{center}
      \begin{tabular}{l l c}
        \hline
	\hline
        Parameter       &        & HD\,180777  \\
	\hline
        Spectral Type   &        & A9V \\
        $v\sin{i}$      & [km\,s$^{\rm -1}$] & 50  \\
        V               &        & 5.11 \\
        B-V             &        & 0.308 \\
        $\pi$           & [mas]  & 36.6 \\
        Distance        & [pc]   & 27.3 \\
        $M_V$           &        & 2.93 \\
        $T_{eff}$       & [K]    & 7250  \\
        log $g$         &        & 4.34  \\
        $M_1$         & [M$_\odot$] &1.7 \\
        \hline
      \end{tabular}
    \end{center}
  \end{table}

  \begin{figure}[t!]
    \centering
    \includegraphics[bb= 70 241 502 709,width=0.8\hsize]{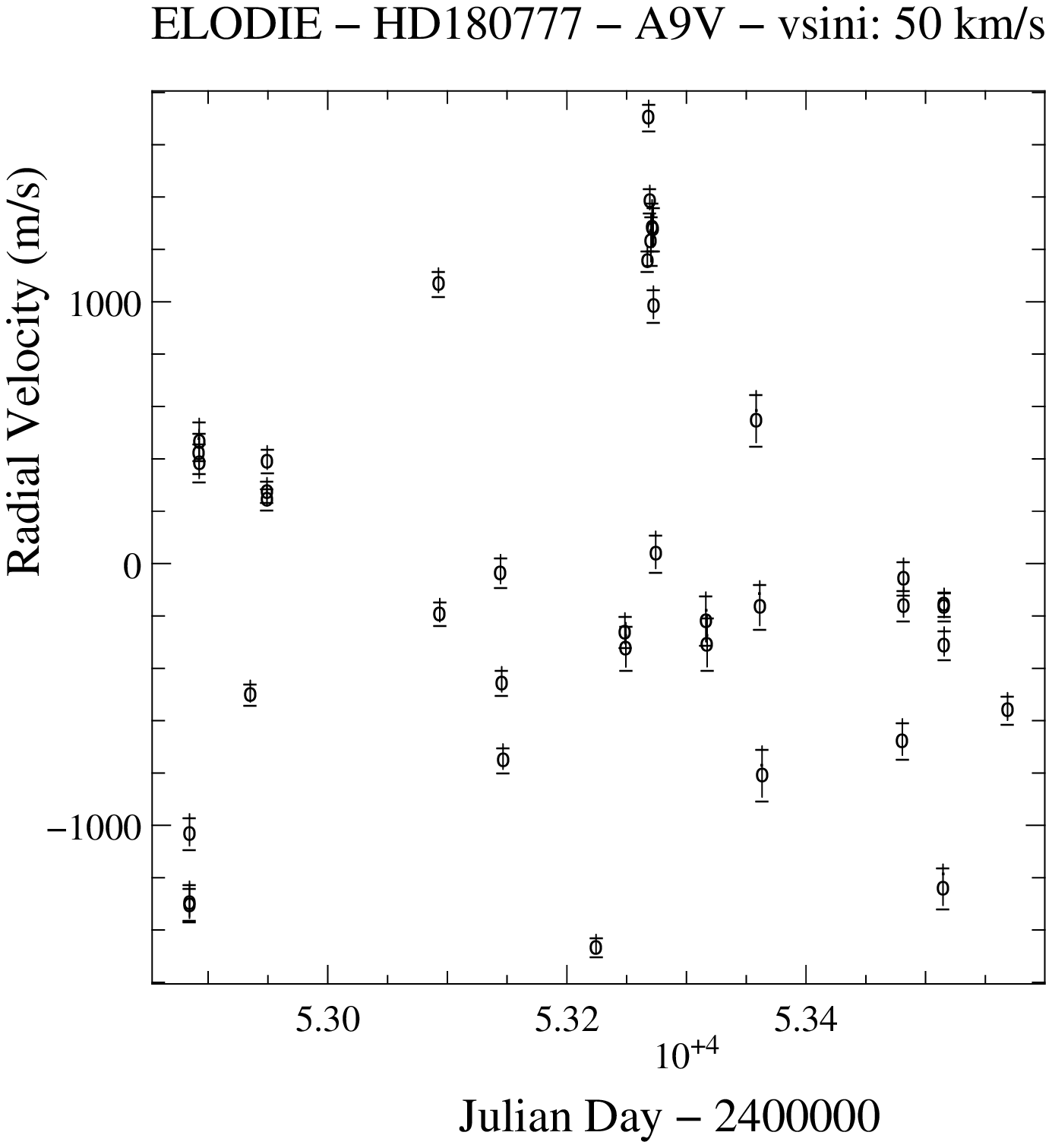}
    \includegraphics[width=0.7\hsize]{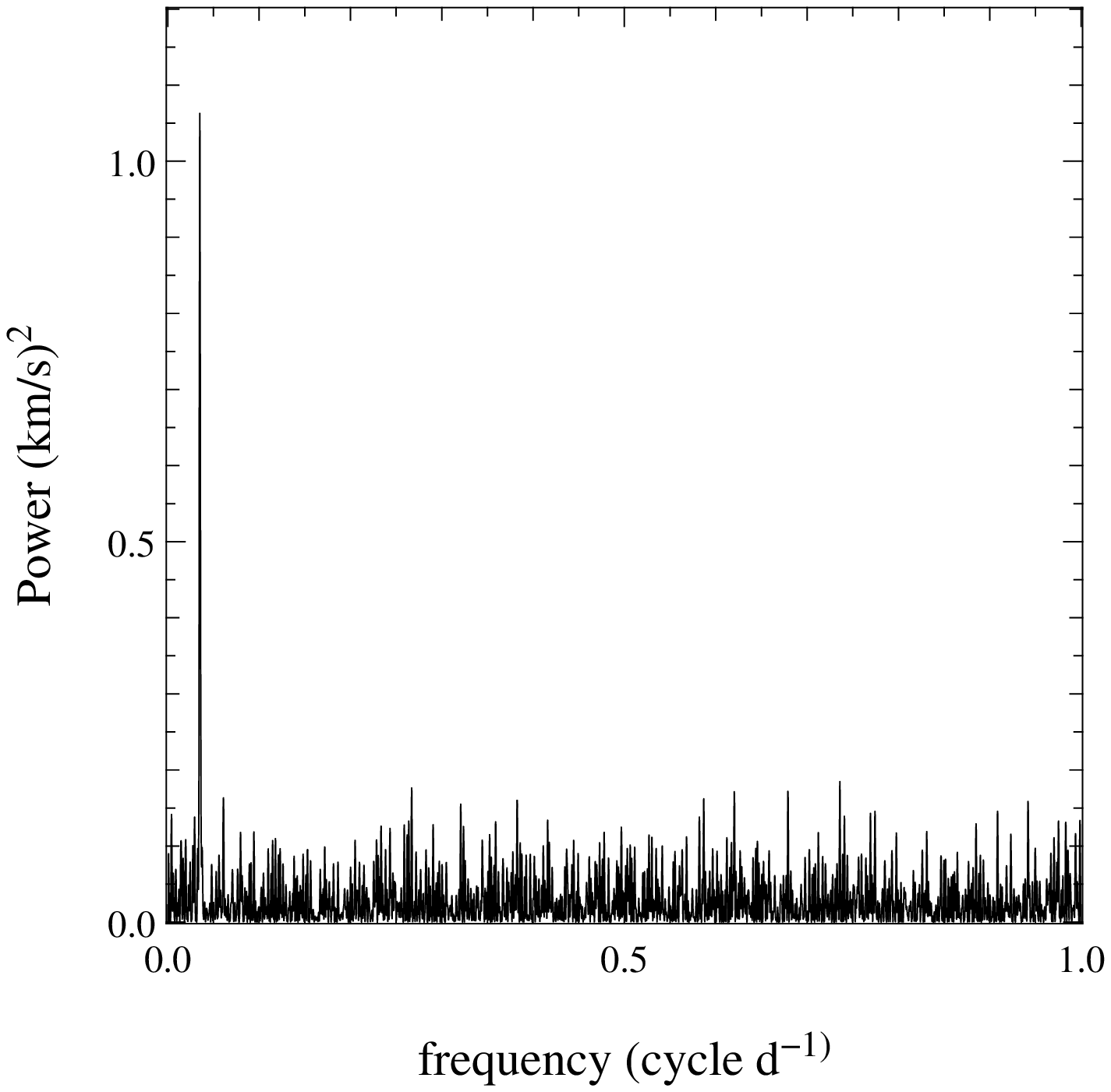}
    \includegraphics[width=0.68\hsize]{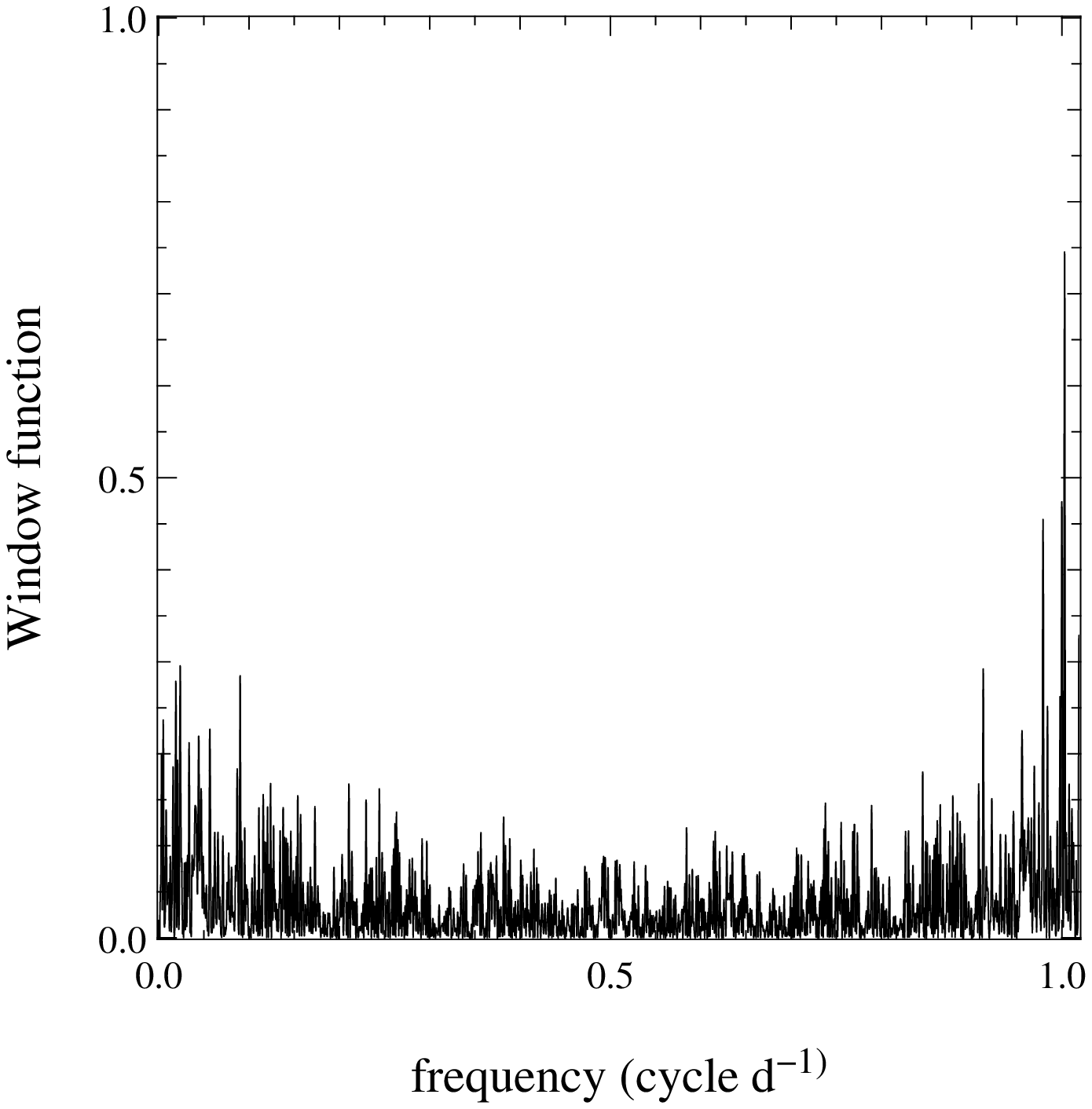}
    \caption{Radial velocities of HD\,180777 obtained with
      \small{ELODIE} (top), the associated periodogram (center) and
      the window function (bottom).}
    \label{180777_vr_perio}
  \end{figure}

  \section{Radial velocity measurements}

  By July 2005, 45 spectra of HD\,180777 were acquired with {\small
  ELODIE}, over 690 days. The wavelength range of the spectra is
  3850-6800\,$\AA$. Six spectra obtained under bad weather conditions
  (with an absorption larger than 2 mag) were discarded.  The typical
  exposure time was 15 min, leading to an $\mathrm{S/N}$ equal to
  $\sim$\,190. The exposures were performed without the
  simultaneous-thorium mode usually used to follow and correct for the
  local astroclimatic drift of the instrument; a wavelengh calibration
  was performed each hour, however, which is largely sufficient to
  correct for the drift in the case of {\small ELODIE} and when the
  radial-velocity photon-noise uncertainties are larger than a few
  dozens m\,s$^{\rm -1}$. In this way, the spectra obtained are not
  polluted by the stronger Thorium lines spread on the CCD, a
  mandatory requirement for our method with A-F spectral type stars
  (see Paper\,I).

  For each spectrum, we selected 34 spectral orders with high
  contrast, covering a wavelength range of 4100-5700\,$\AA$ and
  avoiding orders containing calcium and hydrogen lines or
  contaminated by telluric absorption lines.  Assuming that the
  spectra are translated (stretching in the wavelength space) from one
  to the other, the radial velocities were measured using the method
  described in \cite{Chelli00} and Paper\,I. They are displayed in
  Fig.\,\ref{180777_vr_perio} (top). The uncertainty of 64\,m\,s$^{\rm
  -1}$ on average is consistent with the value of 70\,m\,s$^{\rm -1}$
  obtained from simulations in Paper\,I by applying the relation
  between the radial velocity uncertainties and $v\sin{i}$ to
  HD\,180777, with $\mathrm{S/N}$ values equal to 190.  The observed
  radial velocities are found to be variable with a much larger
  amplitude than the uncertainties.

  The periodogram of the radial velocities is displayed in
  Fig.\,\ref{180777_vr_perio} (center). We used the CLEAN algorithm
  (\cite{Roberts87}) in order to remove the aliases linked with the
  temporal sampling of the data (this algorithm iteratively
  deconvolves the window function from the initial ``dirty''
  spectrum). We used only one iteration here, with a gain loop value
  of one. A clear peak appears at a period of 28.4 days. It is not a
  sampling effect, given the window function
  (Fig.\,\ref{180777_vr_perio}, bottom). Figure \ref{hd180777_dvft}
  shows the radial velocities phased with this period. It is
  consistent with this periodicity in the radial velocity
  variations. In addition, we calculated the false-alarm probability
  of this signal. To do so, we performed a Fisher randomization test
  (\cite{Nemec85}): the radial velocities are shuffled randomly with
  the same time-series as observations, then the periodogram is
  calculated; this is repeated many times (50\,000 here), and the
  number of periodograms containing a power higher than the one for
  the 28.4 days signal in the measured radial velocities is
  stored. The false-alarm probability found this way is lower than
  2.$10^{-5}$, confirming the significance of this 28.4 days signal.

  \section{A brown dwarf around HD\,180777}

  \subsection{Main variation: orbital parameters}

\begin{figure}[t!]
    \centering
    \includegraphics[width=1.0\hsize, height=1.0\hsize]{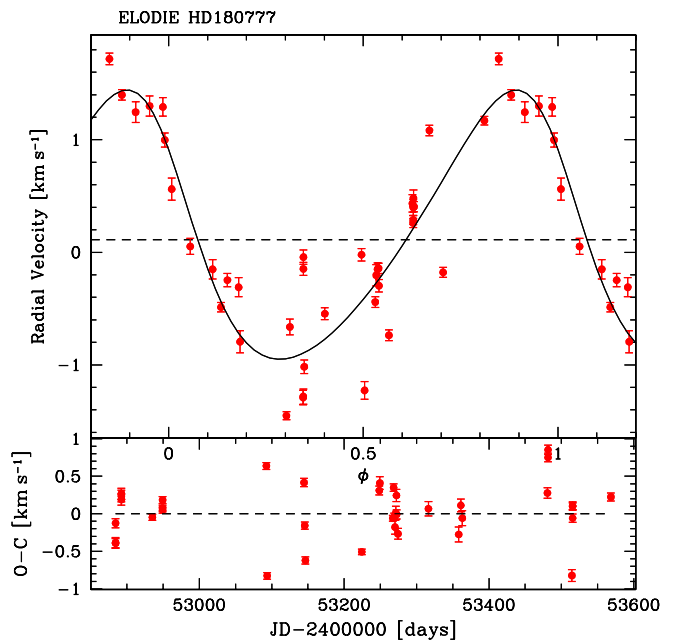}
    \caption{ Top: {\small ELODIE} radial velocities and orbital solutions for
    HD\,180777. Bottom: Residuals to the fitted orbital solution.}
    \label{hd180777_dvft}
  \end{figure}

  We fit the radial velocities with a Keplerian solution
  (Fig.\,\ref{hd180777_dvft}). The orbital parameters derived from the
  best solution are given in Table\,\ref{Table_hd180777_rvpar}.  The
  amplitude is 1.2\,km\,s$^{\rm -1}$, which is consistent with the
  value of the peak in the previous periodogram.  The orbital period
  is 28.4\,days, and the eccentricity is 0.2. Assuming a primary mass
  of 1.7\,M$_\odot$, the companion mass falls in the brown-dwarf
  domain, with a minimum mass of 25\,M$_{\rm{Jup}}$.  Note that a
  value of 0.4 for $\sin{i}$ (see Sect.~2) would result in a true mass
  value of 62\,M$_{\rm{Jup}}$; we even cannot completely exclude the
  case of a low mass M dwarf.  The separation between this candidate
  brown dwarf and the star is 0.2\,AU.  The dispersion of the
  residuals is 394\,m\,s$^{\rm -1}$. They are much larger than
  uncertainties (also clearly seen on the individual residuals,
  Fig.\,\ref{hd180777_dvft}), suggesting another source of radial
  velocity variations. This source should not be a companion since
  there is no satisfactory Keplerian fit to these residuals
  considering the case of only one companion (in addition to the above
  brown dwarf).

  \begin{table}[t!]
    \caption{{\small ELODIE} best orbital solution for HD\,180777.}
    \label{Table_hd180777_rvpar}
    \begin{center}
      \begin{tabular}{l l c}
	\hline
        \hline
        Parameter     &                          & HD\,180777 \\
        \hline
        P             & [days]                   & $ 28.44  \pm $ 0.01 \\
        T             & [JD-2400000]             & $ 53244.3\pm $ 0.3 \\
        e             &                          & $ 0.20   \pm 0.01 $ \\
        $\gamma$      & [km\,s$^{\rm -1}$]       & $ 0.11   \pm $ 0.01\\
        $\omega$      & [deg]                    & $ 56     \pm 4$    \\
        K             & [km\,s$^{\rm -1}$]       & $ 1.20   \pm 0.02$ \\
        N$_{meas}$    &                          & 39     \\
        $\sigma(O-C)$ & [km\,s$^{\rm -1}$]       & 0.394  \\
        \hline
        $a_1\sin{i}$  & [$10^{-3}$~AU]           & 3.1   \\
        f(m)          & [$10^{-6} \rm{M}_{\odot}$] & 4.7   \\
        $M_1$         & [M$_{\odot}$]            & 1.7   \\
        $m_2\sin{i}$  & [M$_{\rm{Jup}}$]              & 25    \\
        $a$           & [AU]                     & 0.22  \\
        \hline
      \end{tabular}
    \end{center}
  \end{table}

  \begin{figure}[t!]
    \centering
    \includegraphics[width=0.47\hsize]{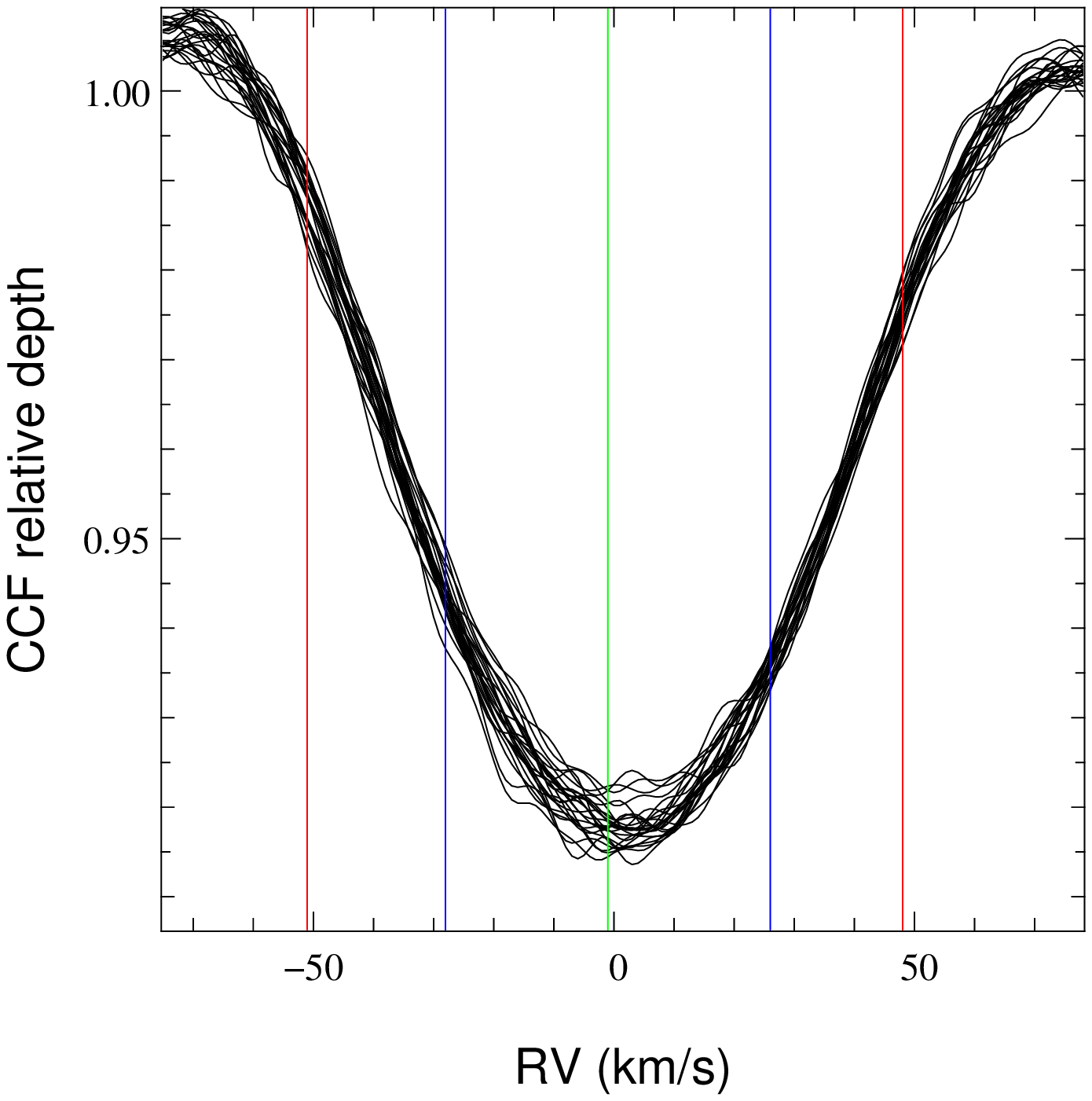}
    \includegraphics[width=0.51\hsize]{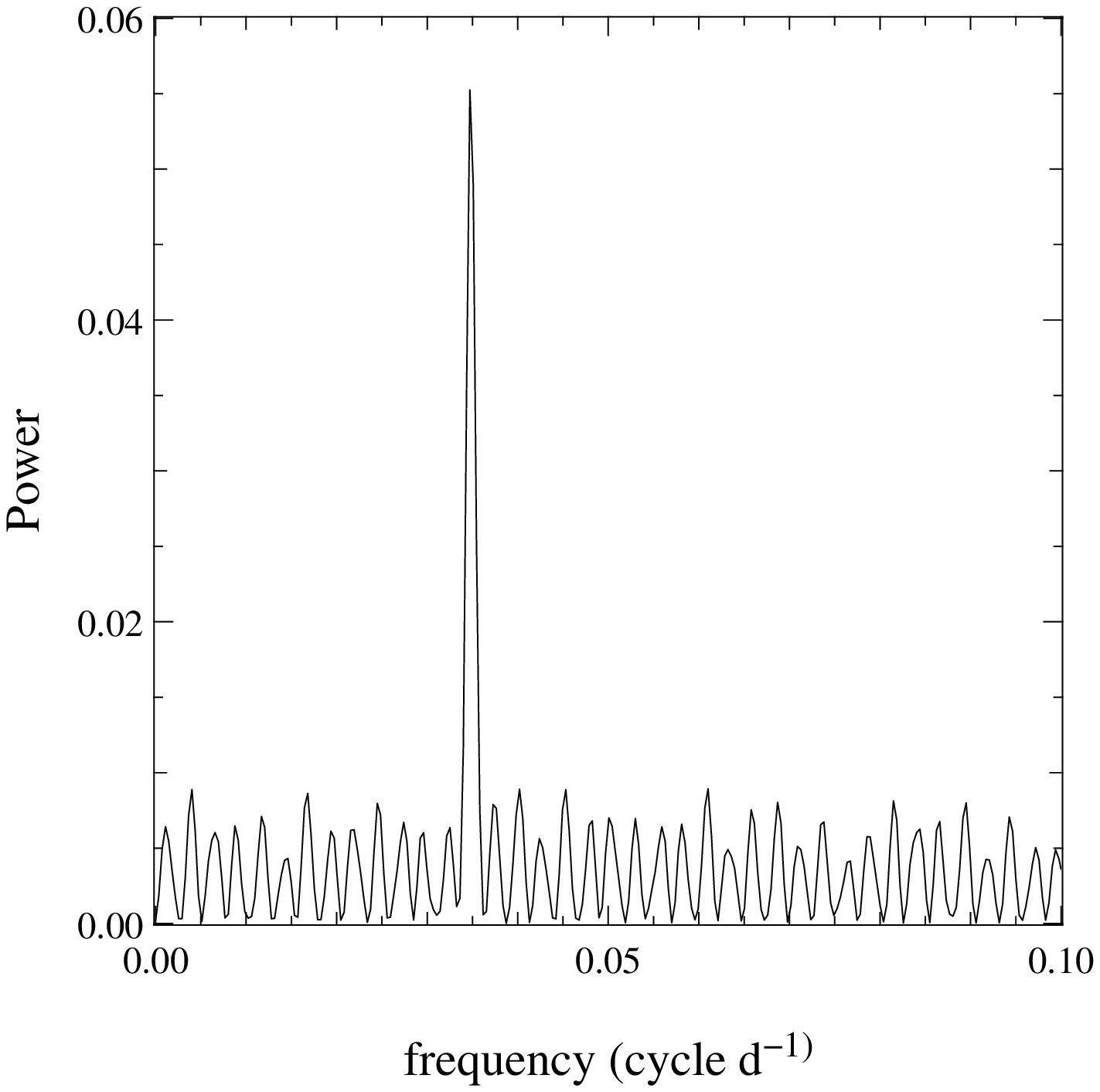}
    \includegraphics[width=0.47\hsize]{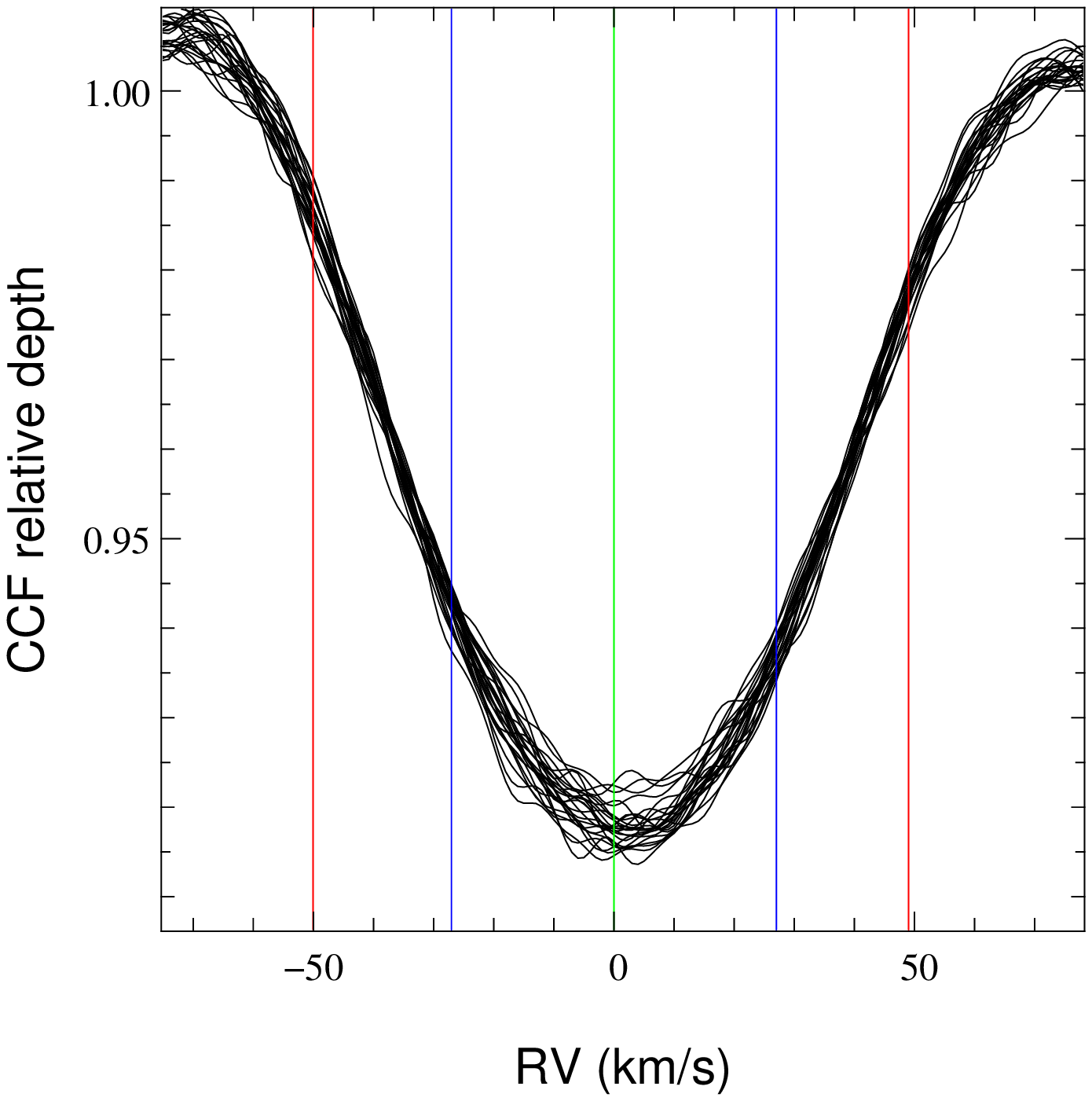}
    \includegraphics[width=0.505\hsize]{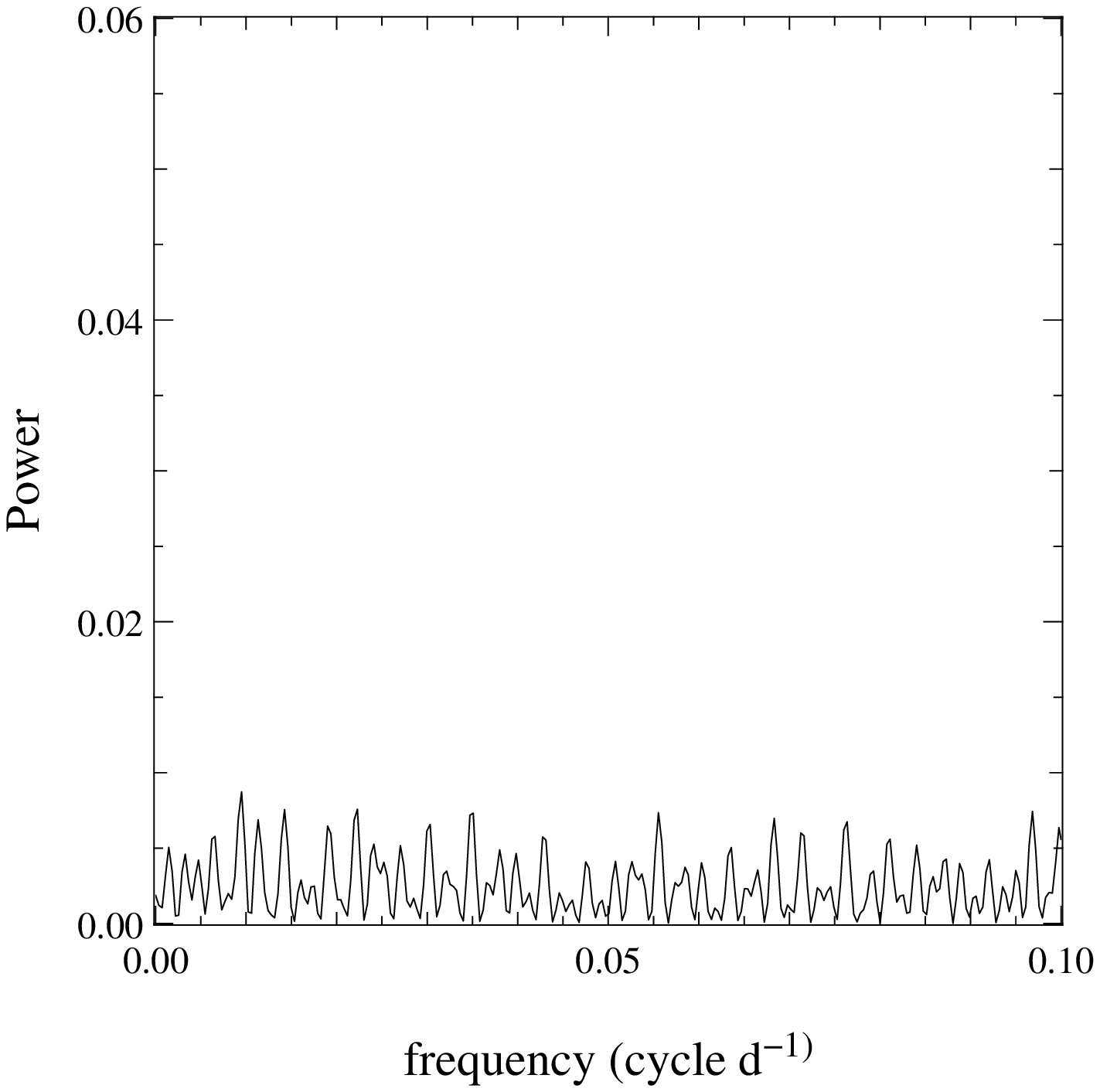}
    \caption{Left: cross-correlation functions of HD\,180777, before (top)
      and after (bottom) correction from the Keplerian motion.
      Right: their corresponding summed periodograms (see text), with the same
      scale in the two cases.}
    \label{180777_CCF}
  \end{figure}

  \subsection{Checking the translation of the spectra}

  We check here that the above periodic signal is only due to a
  periodic translation of the spectra, without simultaneous change in
  the shape of the lines that would correspond to variations that are
  intrinsic to the star. To do so, we compute the cross-correlation
  function of each spectrum with a binary mask. These correlation
  functions represent the mean line profile of each spectrum, and they
  are displayed in Fig.\,\ref{180777_CCF} (top, left).  Note that this
  is a standard way to measure the radial velocities for solar-type
  stars. The correlation functions show a dispersion from one to the
  other, potentialy due in part to a translation, visible in
  particular in the zones of large slope. The dispersion of the
  positions of Gaussian fits made considering only these zones is 870
  m\,s$^{\rm -1}$.

  In order to detect periodic variations in the lines of the spectra,
  we calculated the temporal periodogram corresponding to each point
  of the cross-correlation functions. We then summed all these
  periodograms, so as to enhance the variations occuring along all the
  cross-correlation functions. This summed periodogram is displayed in
  Fig.\,\ref{180777_CCF} (top, right). A clear peak appears at a
  frequency corresponding to a period of 28.4 days.  The radial
  velocities found considering the center of the Gaussian fits are
  consistent with the ones measured in Sect.~3, given the
  uncertainties.

  We then translated the cross-correlation functions to correct them
  from the orbital solution found in Sect~4.1. The results are
  displayed in Fig.\,\ref{180777_CCF} (bottom, left). The dispersion
  of the positions of Gaussian fitted only to the zones of large slope
  is now 410 m\,s$^{\rm-1}$, half the value found previously. It is
  also consistent with the dispersion of the radial-velocity residuals
  around the Keplerian solution. This is a first check of the reality
  of the translation of the spectra from one to the others.  Moreover,
  the summed periodogram, displayed in Fig.\,\ref{180777_CCF} (bottom,
  right), does not show any peak at the frequency corresponding to a
  period of 28.4 days. This confirms that the cross-correlation
  functions were effectively translated from one to the others, and
  that the fit of the radial velocities is accurate.  Note that a
  correcting translation of the cross-correlation functions made at a
  wrong period and/or a wrong amplitude would produce or enhance a
  peak at the corresponding frequency in the summed periodogram,
  instead of removing it. In the same way, if there was no initial
  translation in the spectra but only changes in the shapes of the
  lines, the peak in the periodogram would not disappear after
  corrected translation of the cross-correlation functions.

\begin{figure}[t!]
    \centering
    \includegraphics[bb= 68 251 522 729,width=0.75\hsize]{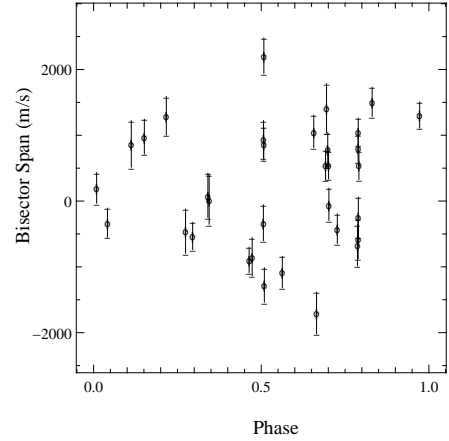}
    \caption{
      Span of the cross-correlation functions phased with a period of
      28.44 days:  no periodic variation with this period.
    }
    \label{180777_span-jj-phas}
  \end{figure}

  We finally check that this periodic translation of the spectra is
  not accompanied by simultaneous changes in the shape of the lines
  with the same period of 28.44
  days. Figure\,\ref{180777_span-jj-phas} shows the span (or inverse
  slope) of the bisectors of the cross-correlation functions, phased
  with this period. They are significantly variable, indicating
  changes in the shape of the lines (see Sect.~6), but there is no
  periodic variation in the spans with a period of 28.44 days. Hence,
  the 28-d signal corresponds only to a shift of the spectra, and not
  to a change in the line shape. The existence of a brown dwarf is the
  best explanation for this signal.

  \section{Ruling out other origins of the main 28-d variations}
  
  \subsection{A single star}

  We checked that HD\,180777 is not a blended double-lined
  spectroscopic binary. In such a case, the FWHM of the
  cross-correlation functions is expected to be linked with their
  depth (anti-correlation). No such correlation is observed for
  HD\,180777.

  \begin{figure}[t!]
    \centering
    \includegraphics[bb= 68 251 522 729,width=0.75\hsize]{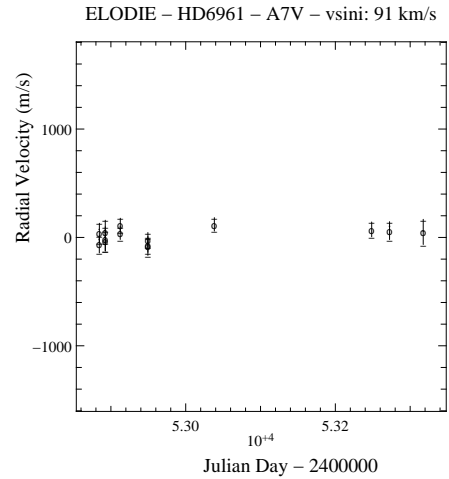}
    \caption{
      {\small ELODIE} radial velocity data for HD\,6961, a
      star constant in radial velocity (dispersion of
      63\,m\,s$^{\rm-1}$), similar and close to HD\,180777.
      The vertical scale is identical to the one in
      Fig.\,\ref{180777_vr_perio}.}
    \label{hd6961_vr}
  \end{figure}
  
  \begin{figure}[t!]
    \centering
    \includegraphics[width=0.65\hsize]{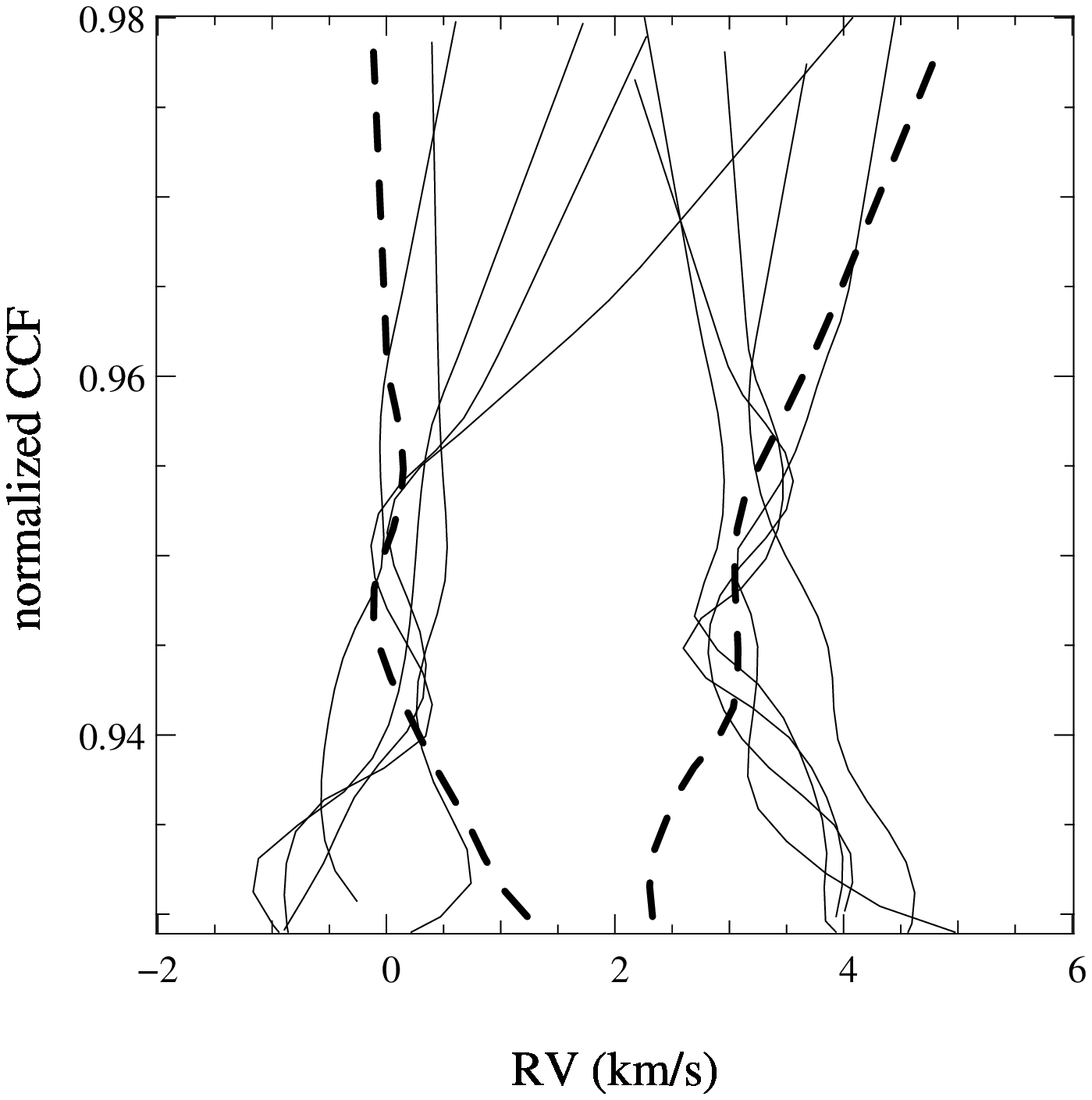}
    \includegraphics[width=0.7\hsize]{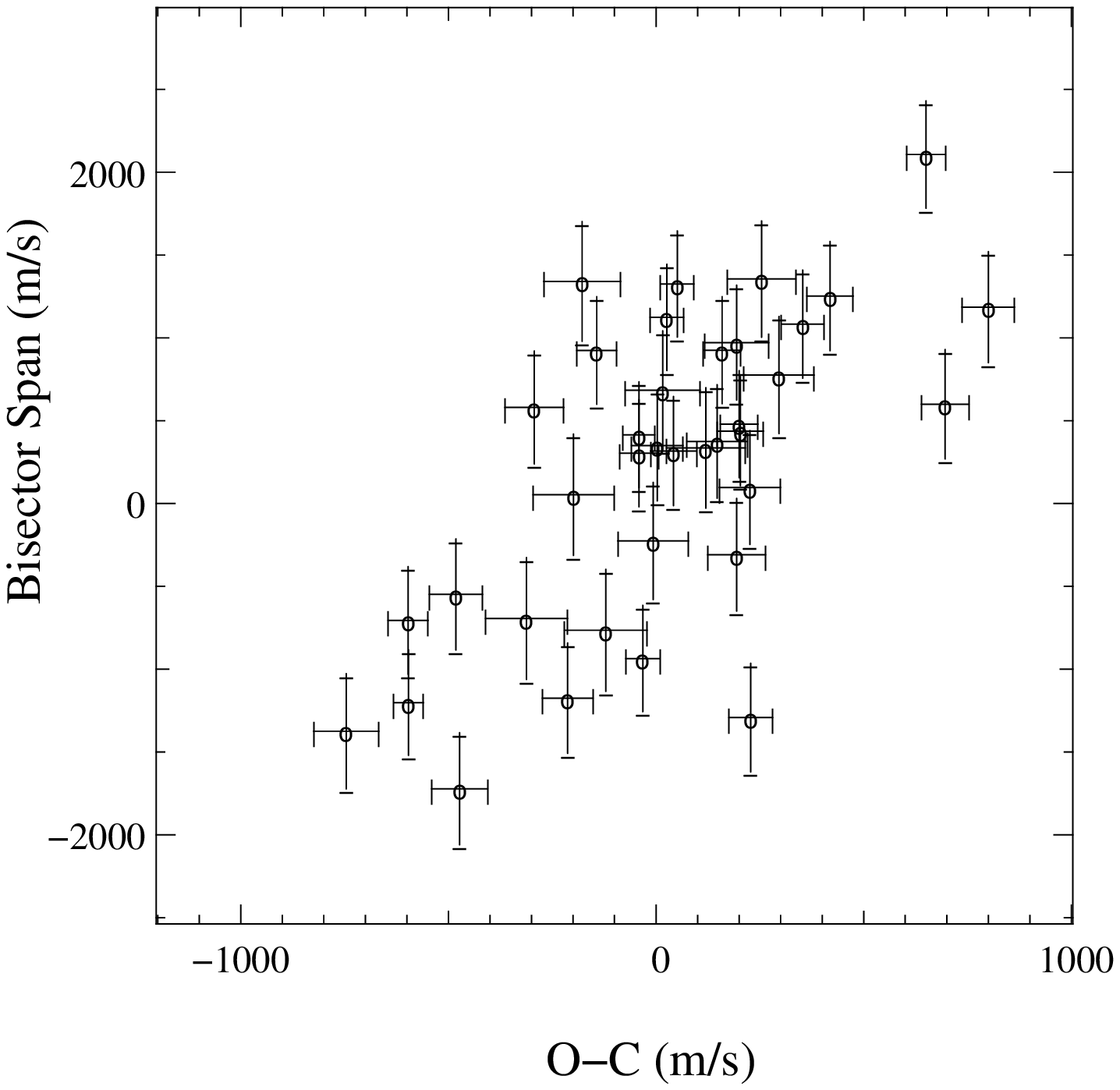}
    \includegraphics[width=0.7\hsize]{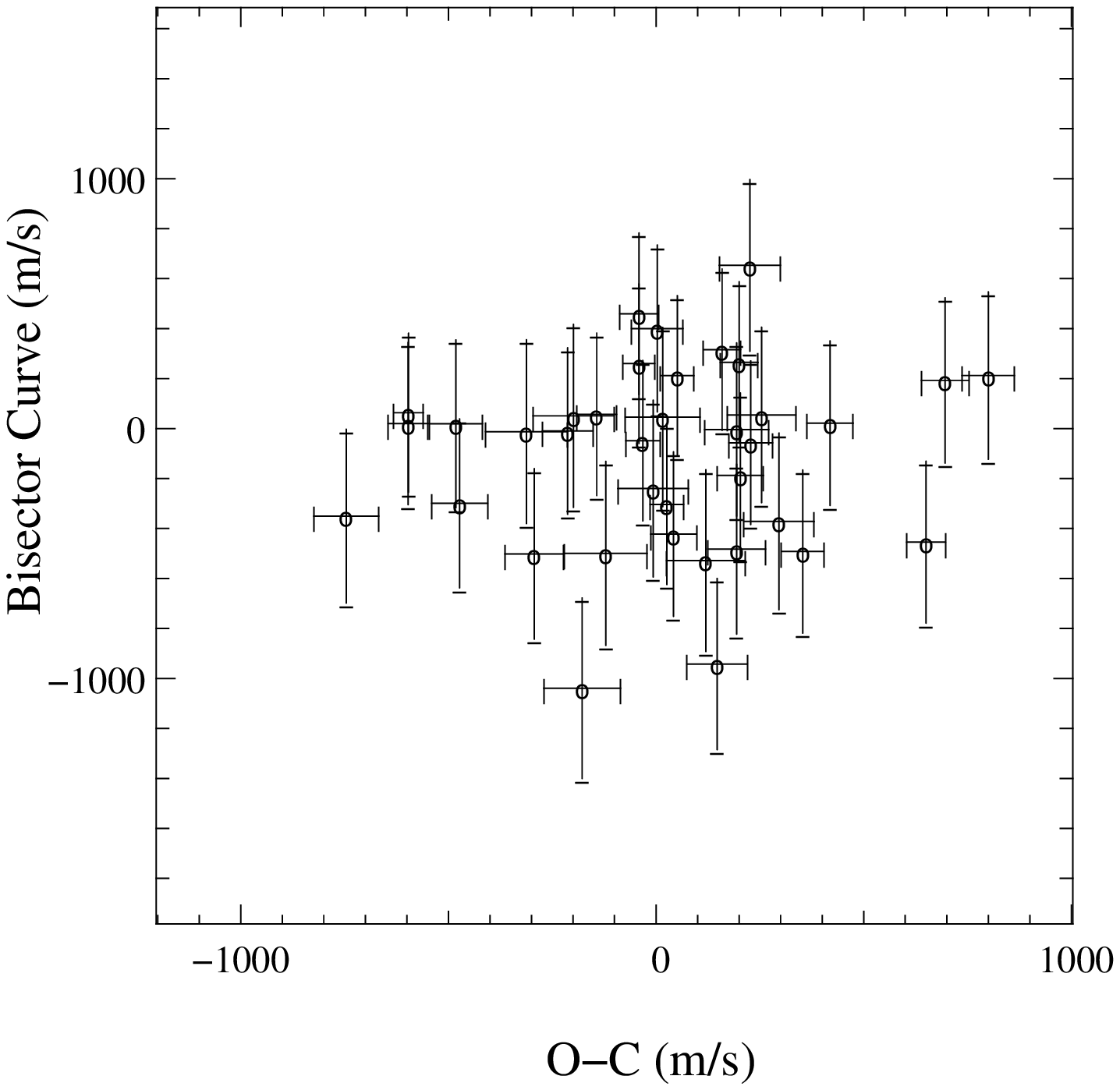}
     \caption{Top: bisectors of the cross-correlation
     functions of HD\,180777; only the bisectors of the spectra giving
     the largest residuals are represented, in 2 sets. Those corresponding to
     large positive (resp. negative) residuals have been translated to
     the left (resp. right), in order to better see the difference
     between them. Dashed lines represent the median
     bisectors of the dual set.
     Center: span of the bisectors as a function of the radial
     velocity residuals.
     Bottom: curvatures of the bisectors as a function of the residuals.
     }
     \label{180777_bis-rec-VR-aj_sel-OC_span}
  \end{figure}
  \subsection{A similar star constant in radial velocity}

  HD\,180777, with a declination of +76 degrees, is located far from
  the ecliptic plane, i.e. always far from the Moon. Moreover, the
  spectral type of the Sun, whose light is reflected by the Moon, is
  very different from the one of HD\,180777; cut frequencies applied
  during the radial velocity computation (Paper\,I) thus eliminate a
  potential contamination of the spectra.
 
  Still, to rule out any possibility of artifact linked with the
  Moon's orbital motion, we show the radial velocities of a similar
  star in Fig.\,\ref{hd6961_vr} , close to HD\,180777 (declination of
  +55 degrees), but constant within the present level of
  uncertainties: HD\,6961 is an A7V star with
  $v\sin{i}$\,=\,91\,km\,s$^{\rm -1}$. It also belongs to our {\small
  ELODIE} sample.  By July~2005, 15 spectra were gathered for this
  star, with an $\mathrm{S/N}$ equal to 266 on average.  The typical
  uncertainty is 83\,m\,s$^{\rm -1}$, comparable to the one obtained
  for HD\,180777. The observed radial velocity dispersion of 63
  m\,s$^{\rm -1}$ shows that this star is constant over the 440~days
  of the measurements (Fig.\,\ref{hd6961_vr}).

  \begin{figure}[t!]
    \centering
    \includegraphics[width=0.61\columnwidth]{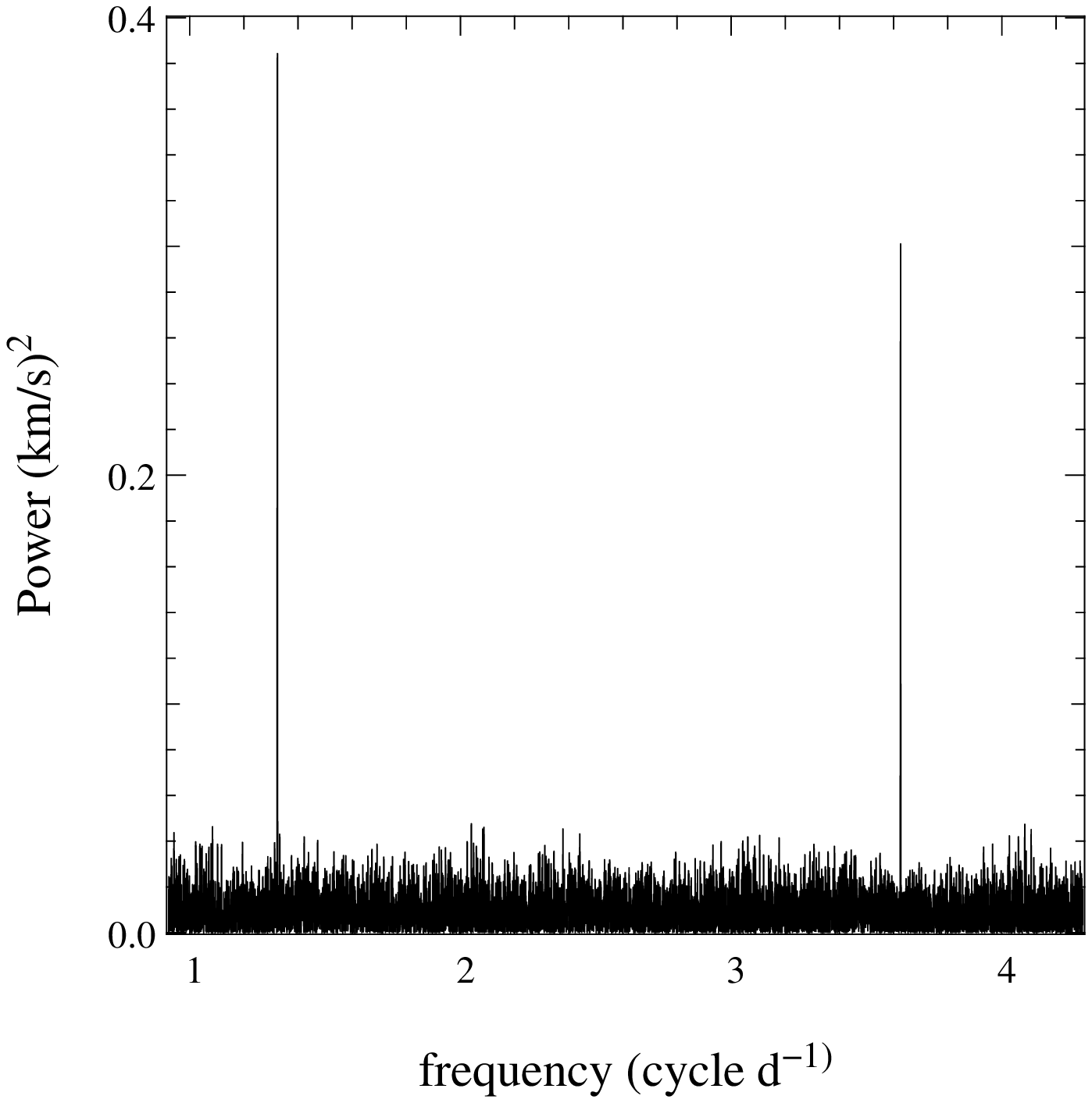}
    \includegraphics[width=0.61\columnwidth]{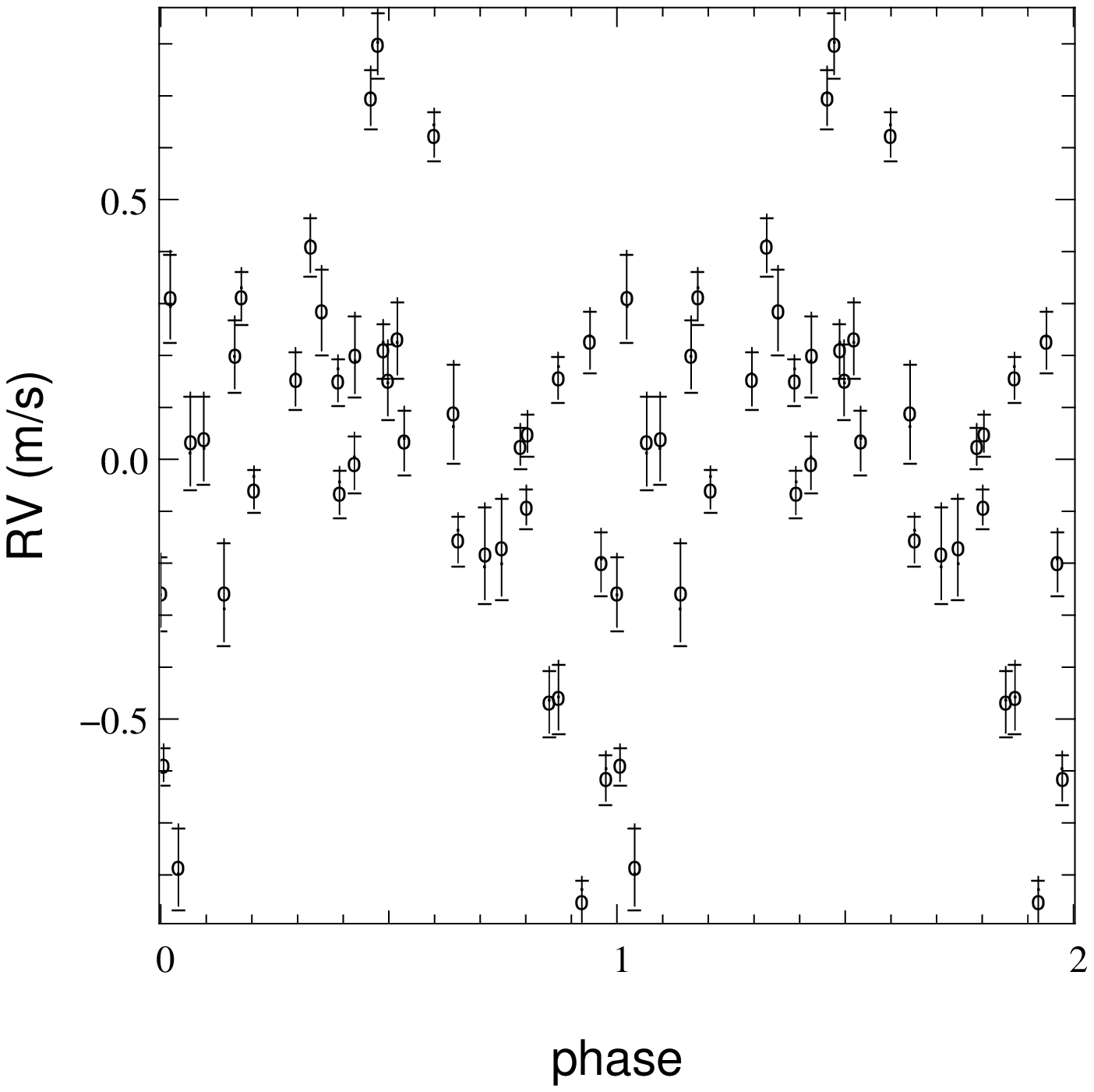}
    \includegraphics[width=0.61\columnwidth]{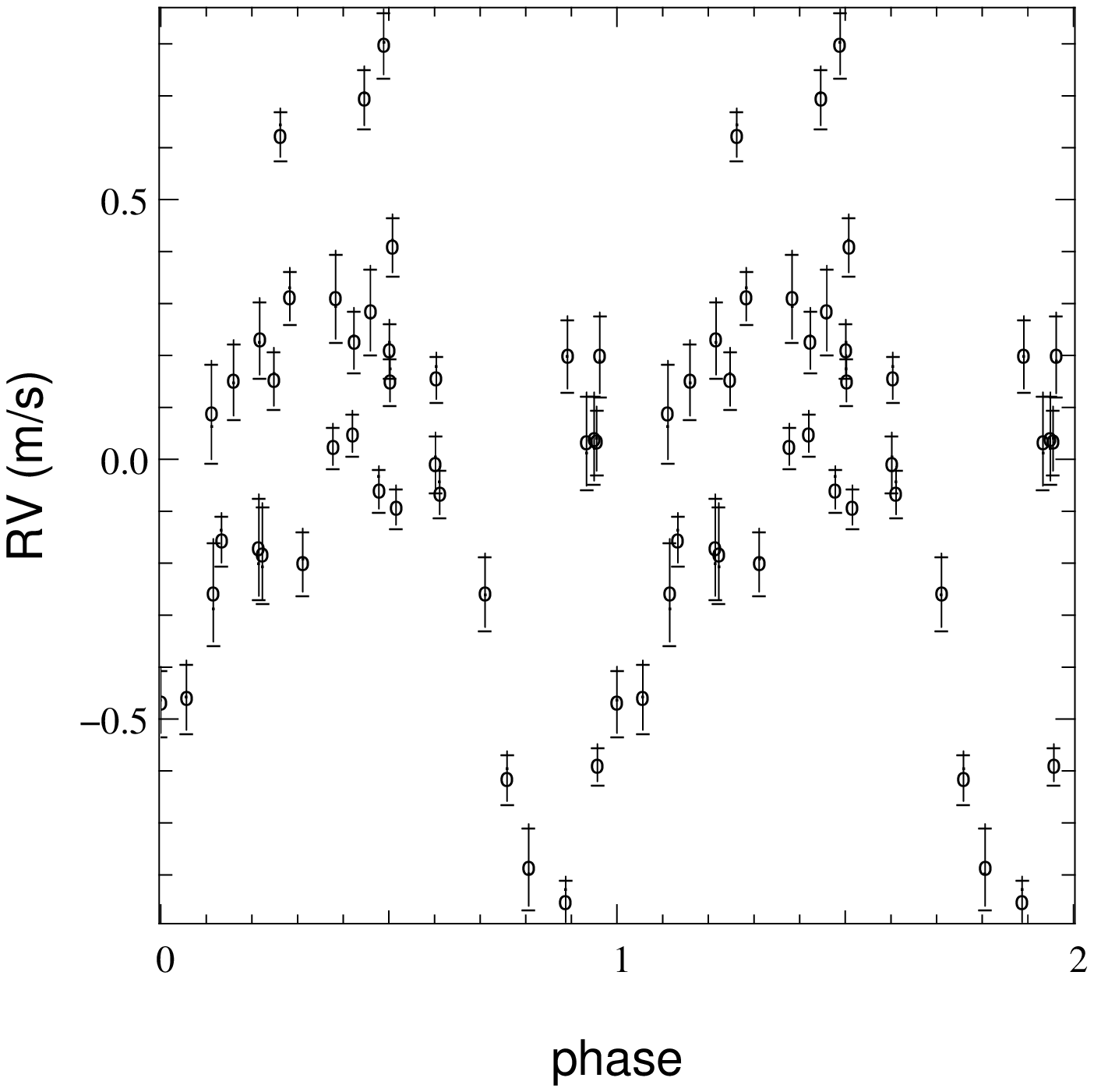}
    \caption{High frequency periodograms of the radial velocities
    obtained on HD\,180777 (top) and the phasing
    of the radial velocities to the corresponding periods (bottom).
    }
    \label{El_dvft-perio_HF}
  \end{figure}
  \section{Radial-velocity residuals: pulsations}

  Considering the cross-correlation functions again, we can
  investigate whether the large radial-velocity residuals observed
  around the orbital solution can be related to the changes in the
  shape of these mean line profiles. We then compute the bisector of
  the cross-correlation functions
  (Fig.\,\ref{180777_bis-rec-VR-aj_sel-OC_span}, top). A first step
  consists in calculating their span (or inverse slope), and to look
  for a correlation between the spans and the radial-velocity
  residuals. Figure \ref{180777_bis-rec-VR-aj_sel-OC_span} (center)
  shows the bisector spans as a function of these residuals: they seem
  to be linearly correlated, with a slope value close to 2. The
  changes in the shape of the lines thus appear to be responsible for
  at least a part of the radial-velocity residuals considered. Cool
  spots linked with magnetic activity are unlikely because in this
  case, the slope is negative (\cite{Queloz2001a}). Hot spots could be
  responsible for this correlation and cannot be excluded, although
  they are unlikely. Besides, the presence of a stellar binary system
  can produce this sort of linear correlation with a positive slope
  (\cite{Santos02}), but we checked (see Sect.~5) that it is not the
  case here, at least if the flux received from the two stars is
  similar. A remaining explanation is the presence of low order
  pulsations (\cite{Hatzes96}).

  As a second step, we calculated the curvature of the bisectors of
  the considered cross-correlation functions, as it is also shown to
  be useful for characterizing the pulsations (\cite{Hatzes96}). The
  results are displayed on
  Fig.\,\ref{180777_bis-rec-VR-aj_sel-OC_span} (bottom) and do not
  show any correlation with the radial-velocity residuals.  The
  bisectors thus mainly change with regard to their span; for a given
  variation of the span, we then expect around half this variation in
  the radial velocities (averaging effect). The slope value close to 2
  found above between the bisector spans and the radial-velocity
  residuals then shows that the changes in the shape of the lines
  fully explain the dispersion of these residuals.

  Even if our temporal sampling does not allow for a detailed analysis
  of short period variations, we are still able to enhance two
  frequencies characteristic of pulsations, at
  1.324$\pm0.001$\,cycle\,d$^{-1}$ (period of 18.1\,h) and
  3.626$\pm0.001$\,cycle\,d$^{-1}$ (period of 6.6\,h)
  (Fig\,\ref{El_dvft-perio_HF}, top). The false-alarm probabilities of
  these peaks are 0.34\% and 8.4\%, respectively, indicating a high
  significance for the 1.324\,cycle\,d$^{-1}$ signal, but a lower one
  for the 3.626\,cycle\,d$^{-1}$ signal. The phasing of the radial
  velocities to the corresponding periods is consistent with the
  existence of these signals (see Fig\,\ref{El_dvft-perio_HF},
  bottom), as well as the amplitudes of these radial-velocity
  variations (typically 200\,m\,s$^{\rm -1}$). A fit of the radial
  velocities with the superposition of two sinusoids with periods
  fixed to the above values leads to a decrease in the radial-velocity
  dispersion from 394 to 239\,m\,s$^{\rm -1}$, which is still well
  above the uncertainties (64\,m\,s$^{\rm -1}$ on average). This
  convergence happens for values of the amplitude of 239 and
  234\,m\,s$^{\rm -1}$, respectively.  We are not able to detect other
  high frequencies, probably because of our temporal sampling, which
  is not really adapted to the search for high frequency variations.

  The radial-velocity residuals observed around the orbital solution
  are thus very probably explained by changes in the shape of the
  lines created by pulsations of the star. The presence of pulsations
  in the case of HD\,180777 is not surprising, as this star belongs to
  the range of B-V where the instability strip intersects with the
  main sequence, and where we find the pulsating $\delta$~Scuti
  (\cite{Handler02}, \cite{Breger00}) and $\gamma$~Dor stars
  (\cite{Mathias04}, \cite{Handler02}).  As the frequencies of the
  variations are lower than 0.25 cycle\,d$^{\rm -1}$ (periods larger
  than 6.5\,h), HD\,180777 should probably belong to the pulsating
  $\gamma$~Dor stars.  As these stars undergo non-radial pulsations
  resulting in multi-periodic radial-velocity variations with an
  amplitude up to several km\,s$^{\rm -1}$, the level of 400
  m\,s$^{\rm -1}$ found here for the radial-velocity residuals appears
  to be common.

  We checked that the peak at 28.4 days is not an alias of these
  higher frequency signals. To do so, we first fitted the initial
  radial velocities with a double sinusoid with periods corresponding
  to the two high frequencies found above. The periodogram of the
  residuals obtained this way still shows a strong peak (with the same
  amplitude as in Fig.\,\ref{180777_vr_perio}, bottom), at a frequency
  corresponding to the same period of 28.4 days. Hence, the signal at
  28.4 days is not an alias of these high frequency signals.

  \section{Concluding remarks}

  We have presented here the first detection of a brown dwarf around
  one of the objects surveyed in our {\small ELODIE} programme,
  HD\,180777, an A9V star with $v\sin{i}$\,=\,50\,km\,$s ^{\rm -1}$.
  This detection is an example of disentangling the presence of a low
  mass companion from the existence of pulsations. The best Keplerian
  solution derived from the radial-velocity measurements leads to a
  minimum mass of 25\,M$_{\rm{Jup}}$ (the true mass could be
  significantly higher) and a period of 28.4~days, hence a separation
  of 0.2~AU.

  It is interesting to note that the detected companion falls in the
  middle of the brown-dwarf desert observed for G-M dwarf primaries.
  For the first time, we are able to probe the mass-ratio of binaries
  with A-type dwarf primaries down to very small mass-ratios.
  
  This result is another step toward extending the study of planet and
  brown-dwarf formation processes to stars earlier than F7. This is
  fundamental to a global understanding of the most interesting
  planetary formation mechanisms involved. In particular, the proposed
  idea that the planet formation process could scale with the primary
  mass is very interesting. Studies on lower mass stars (\cite{Lin05})
  show such a trend between the masses of the primaries and the
  companions. This idea is also consistent with the present detection
  of a brown dwarf around an A9V star. In such a picture, could brown
  dwarfs be formed in the same way as planets ?
 
  \begin{acknowledgements}

    \vspace{1cm}

    We acknowledge support from the French CNRS. We are grateful to
    the Observatoire de Haute-Provence (OHP) and to the Programme
    National de Plan\'etologie (PNP, INSU), for the time allocation,
    and to their technical staff.  
    These results have made use of the SIMBAD database, operated at CDS,
    Strasbourg, France.

  \end{acknowledgements}


\end{document}